\documentclass[12pt,preprint]{aastex}
\begin{document}

\title{Helioseismological Implications of Recent Solar Abundance Determinations}
\author{John N. Bahcall}
 \affil{Institute for Advanced Study, Einstein Drive,
Princeton, NJ
  08540}
\author{Sarbani Basu}
\affil{Department of Astronomy, Yale University, New Haven, CT 06520-8101}
\author{Marc Pinsonneault}
\affil{Department of Astronomy, Ohio State University, Columbus, OH
  43210}
\and
\author{Aldo  M. Serenelli}
 \affil{Institute for Advanced Study, Einstein Drive,
Princeton, NJ
  08540}

\begin{abstract}
We show that standard solar models  are in good agreement with the
helioseismologically determined sound speed and density as a
function of solar radius, the depth of the convective zone, and
the surface helium abundance, as long as those models do not
incorporate the most recent heavy element abundance
determinations. However, sophisticated new analyses of the solar
atmosphere infer lower abundances of the lighter metals (like C,
N, O, Ne, and Ar) than the previously widely used  surface
abundances. We show that solar models that include the lower heavy
element abundances  disagree with the solar profiles of sound
speed and density as well as the depth of the convective zone and
the helium abundance. The disagreements for models with the new
abundances range from factors of several to many times the quoted
uncertainties in the helioseismological measurements.  The
disagreements are at  temperatures that are too low to affect
significantly  solar neutrino emission. If errors in the
calculated OPAL opacities are solely responsible for the
disagreements, then the corrections in the opacity must extend
from $2 \times 10^6$K ($R = 0.7R_\odot$) to $5\times 10^6$K ($R =
0.4 R_\odot$), with opacity increases of order 10\%.
\end{abstract}


\section{INTRODUCTION}
\label{sec:intro}

Why are precision tests of solar models important?  The Sun is a
laboratory in which the predictions of stellar evolution theory
can be compared with uniquely accurate and detailed measurements.
Stellar evolution calculations are used throughout astronomy to
classify, date, and interpret the spectra of individual stars and
of galaxies. Comparisons, discussed  in this paper, between
helioseismological measurements and solar model calculations
suggest that at least one of the ingredients of stellar evolution
calculations is not known as precisely as previously believed. We
shall see that there are reasons for questioning the accuracy of
the  most sophisticated and detailed determinations of stellar
abundances, the recent measurements of the solar heavy element
abundances. Alternatively, unexpectedly large changes could be
required in the radiative opacity. However, we shall also see that
the disagreement between helioseismological measurements and solar
model predictions (with the new metal abundances) occur at
relatively low temperatures and therefore do not affect
significantly the predicted solar neutrino fluxes.

 Helioseismology provides
sensitive and powerful tests of the theory of stellar evolution.
In addition to measuring the depth of the solar surface convection
zone and the surface helium abundance, inversions of seismic data
are used to measure to high precision the speed of sound as a
function of depth in the star for almost the entire solar
interior. The density distribution can also be determined,
although with an order of magnitude less precision than for the
sound speed.

 A number of investigators have made comparisons of seismic data with
solar models and  have confirmed that the standard solar mixture
of Grevesse \& Noels (1993)\nocite{grevesse93}
 and the updated mixture of Grevesse \&
Sauval (1998)\nocite{oldcomp} yield solar models in good agreement
with the data (e.g., Bahcall \& Pinsonneault 1995\nocite{BP95};
Christensen-Dalsgaard, et al. 1996\nocite{jcd96}; Bahcall,
Pinsonneault \& Basu 2001; Christen-Dalsgaard 2002\nocite{jcd02};
Couvidat, Turck-Chieze, \& Kosovichev 2003\nocite{couvidat03};
Sackmann \& Boothroyd 2003\nocite{sackmann03};
 Richard, Th\'eado, \& Vauclair 2004\nocite{richard04}, and references therein.) As
early as 1988, Bahcall \& Ulrich (1988)\nocite{bahcallulrich1988}
showed that detailed solar models computed with the accurate
physics and the numerical precision required for solar neutrino
predictions yielded results in good agreement with the
then-available helioseismological data.

 In a series of papers that preceded the epochal and definitive
  measurements of the SNO  and Super-Kamiokande solar neutrino experiments
  ( Ahmad, et al. 2001\nocite{ahmad01}; Fukuda et
  al. 2001\nocite{fukuda01}), we
  showed that the excellent agreement between the computed sound
  speeds in precise standard solar models and the precise
  helioseismological inversions (differences $< 0.1$\% rms throughout
  the solar interior)implied that new physics was required to solve
  the solar neutrino problem (Bahcall, Pinsonneault,
  Basu, \& Christensen-Dalsgaard 1997\nocite{bpbc97}; Bahcall,
Basu, \& Pinsonneault 1998\nocite{bbp98}; Bahcall, Pinsonneault, \&
  Basu 2001\nocite{BP00}; Bahcall 2001\nocite{bahcall01}).

New and more powerful analyses of the surface chemical composition
of the Sun have recently become available. These new analyses use
three-dimensional atmospheric models, take account of hydrodynamic
effects, and pay special attention to uncertainties in the atomic
data and the observed spectra.  Lower mass fractions have been
obtained in this way for C, N, O, Ne, and Ar (Asplund et al.~2000;
\nocite{asplundetal00} Asplund 2000;\nocite{asplund00} Allende
Prieto, Lambert, \& Asplund 2001;\nocite{allende01} Allende
Prieto, Lambert, \& Asplund 2002;\nocite{allende02} Asplund et
al.~2004;\nocite{asplund04} ). The new abundance determinations,
together with the previous best-estimates for other solar surface
abundances~(Grevesse \& Sauval 1998)\nocite{oldcomp}, imply $Z/X =
0.0176$, much less than the previous value of $Z/X = 0.0229$
(Grevesse \& Sauval 1998)\nocite{oldcomp}. In fact, the recent
estimates for the C, N, and O mass fractions are lower than all
the abundance measurements we have used in the precision solar
models in this series going back to 1971 (see, e.g., Table~II of
Bahcall and \& Pinsonneault 1995).

Despite the great improvement in the techniques now used to
determine the new element abundances, the new abundances cause the
depth calculated for the solar convective zone with the aid of a
standard solar model, $R_{\rm CZ} = 0.726 R_\odot$ (Bahcall \&
Pinsonneault 2004\nocite{BP04}; Bahcall, Serenelli, \& Pinsonneault
2004\nocite{BSP04}; Basu
\& Antia 2004)\nocite{basu04}, to be in strong disagreement with
the measured depth,
\begin{equation}
R_{\rm CZ} = 0.713 \pm 0.001 R_\odot \, ,
\label{eq:radiuscz}
\end{equation}
which is determined by helioseismological techniques (
  Kosovichev \& Fedorova 1991;\nocite{kosovichevcz}
Christensen-Dalsgaard, Gough, \& Thompson 1991;\nocite{jcdCZ}
Guzik \& Cox 1993; Basu \& Antia 1997, 2004\nocite{basucz,basu04};
Basu 1998\nocite{basu98}).\nocite{guzikcz} Paradoxically, the
calculated depth of the convective zone obtained using the older
element abundances, $R_{\rm CZ} = 0.714 R_\odot$, agrees with the
helioseismological value (Bahcall, Pinsonneault, \& Basu
2001)\nocite{BP00}. This situation has been described as the
``convective zone problem'' (Bahcall, Serenelli, \& Pinsonneault
2004)\nocite{BSP04}.

Our goal here is to determine the helioseismological implications
of the recent abundance determinations. We  compare the
helioseismologically measured depth of the solar convective zone,
the sound speed and density as a function of radius,  and the
primordial helium abundance with the values that are obtained
using a series of precise solar models.  The solar models
considered here incorporate the most recent and accurate nuclear
and stellar data, including the equation of state and radiative
opacity.

We describe in \S~\ref{sec:models} the  solar models whose
properties are investigated  in the present paper. We then discuss
in \S~\ref{sec:inversions} the helioseismological data and the
inversion technique that we have used to obtain the measured depth
of the convective zone, the sound speeds and density as a function
of radius, and the initial helium abundance. We compare in
\S~\ref{sec:comparison} the properties of our set of solar models
with the solar parameters that are determined by helioseismology.
We summarize and discuss our conclusions in
\S~\ref{sec:conclusions}.

\section{DESCRIPTION OF SOLAR MODELS}
\label{sec:models}

We describe in this section the basic ingredients of six solar
models that we use to assess the helioseismological implications
of the recent heavy element abundance determinations.

The six solar models considered in detail in this paper are listed
below. Models 1 and 2 were originally computed by Bahcall \&
Pinsonneault (2004)\nocite{BP04}; Model 4 was computed by Bahcall,
Pinsonneault, Serenelli (2004).

\begin{description}
\item[(1) BP04:] older element abundances from Grevesse, \&
Sauvall (1998)\nocite{oldcomp},  and best-available values for all
other input parameters (including improved nuclear rates and
equation of state); \item[(2) BP04+:]  the same as BP04 except
that recent lower estimates for heavy element abundances are
incorporated; \item[(3) BP04--EOS96:] the same as BP04 but with
the OPAL 1996 EOS (Rogers, Swenson, \& Iglesias
1996)\nocite{rog96} instead of the OPAL 2001 EOS (Rogers 2001,
Rogers and Nayfonov 2002\cite{eos2001}); \item[(4) BP04+ 21\%:]
the same as BP04+ except that the OPAL radiative opacity is
increased by 21\% near the base of the convective zone; \item[(5)
BP04+ 11\%:] the same as BP04+ except that the OPAL radiative
opacity is increased by 11\% for temperatures ranging from
$2\times 10^6$K to $5\times 10^6$K; and \item[(6) BP00:] our best
previous-generation standard solar model, obtained by Bahcall,
Pinsonneault, \& Basu (2001)\nocite{BP00} with older values of
nuclear reaction data, an older equation of state (OPAL 1996), and
the Grevesse \& Sauval (1998)\nocite{oldcomp} element abundances.
\end{description}
 The code and techniques used in these calculations
have been described in Bahcall \& Pinsonneault (1992,
1995),\nocite{BP92,BP95} Bahcall \& Ulrich (1988),
\nocite{bahcallulrich1988} Bahcall, Pinsonneault, \& Basu
(2001)\nocite{BP00}.

The reader may wonder why we include in this paper the results
from the model BP00, when BP04 has superseded BP00 by
incorporating  more accurate nuclear reaction data and an improved
equation of state. We include results from BP00 as well as BP04 in
order to have some indication of the kind of differences that can
be expected, independent of solar abundance determinations, as
further improvements are made in the input data to solar models.
The differences between values obtained with the BP00 and the BP04
models may be regarded as within the expected range. We shall see
in what follows that the differences in solar model results caused
by adopting the new heavy element abundance determinations are
much larger than the differences between the results obtained with
BP00 and BP04.

We want our investigations to be as precise as possible and our
inferences to be as free as possible from dependence upon the
idiosyncrasies of a particular stellar evolution code. Therefore,
we have recalculated the BP04, BP04+ and BP04+ 21\% solar models
using the Garching stellar evolution code (see, e.g., Schlattl,
Weiss,  \& Ludwig 1997\nocite{garching} and Schlattl
2002\nocite{schlattl02} for details of the code), to which the
nuclear energy generation routine `exportenergy.f'\footnote{The
routine is publicly available at http://www.sns.ias.edu/\~{}jnb.}
has been coupled. The nuclear cross sections adopted are those
used in Bahcall \& Pinsonneault (2004)\nocite{BP04}. The models
were calculated using the latest version  of the  OPAL  equation
of state  (Rogers 2001)\nocite{OPAL2001},  OPAL radiative
opacities (see below for the composition adopted) and element
diffusion for helium and metals (Thoul, Bahcall, \&  Loeb
1994,\nocite{thoul} code available at the URL given in
footnote~1). The mixing length theory for convection has been used
in all the models. The Schwarzschild  criterion was used to
determine  the  location  of the convective boundaries.

We have verified that the Garching stellar evolution code and the
Bahcall-Pinsonneault code (which has its origins in the CalTech,
UCLA, and Yale codes, see Bahcall \& Ulrich 1988; Bahcall \&
Pinsonneault 1992, 1995;\nocite{BP92,BP95} Prather
1976\nocite{prather}; Pinsonneault, Kawaler, Sofia, \& Demarque
1989)\nocite{pinsonneault89}) yield identical results to the
accuracy of interest in all of the investigations considered in
this paper.

The  model,  BP04, which was  calculated assuming  the older
Grevesse \& Sauval (1998) solar surface composition, has a present
surface ratio of heavy element to hydrogen mass fractions of
$Z/X=0.0229$. The model BP04+, which incorporates the new
determinations of the solar heavy element composition,  has a much
lower ratio of heavy elements to hydrogen, $Z/X=0.0176$. Since new
solar abundance determinations are being reported as they come
available, Table~1 of Bahcall, Serenelli, \& Pinsonneault
(2004)\nocite{BSP04} lists the specific element abundances adopted
in computing both BP04 and BP04+.

 The model BP04+  21\% was designed to bring into agreement the
 calculated and the helioseismologically measured depths of the
 convective zone using a solar model that incorporates the recent
 heavy element abundance determinations.  Bahcall, Serenelli, \&
 Pinsonneault (2004)\nocite{BSP04} showed that a local 21\% increase in
 the tabulated OPAL radiative opacity near the base of the convective
 envelope will produce a model with the base of its convective zone at
 $R_{\rm CZ} = 0.713 R_\odot$, in essentially perfect agreement with
 the measured value for the depth of the convective zone. The factor by which the
 opacity was increased is  similar to the factor needed by Basu \& Antia (2004)
to construct solar envelope models with the new heavy element
abundances that have the same convection zone depth, helium
abundance, and density profile as the Sun.  Very
 recently, Seaton \& Badnell (2004)\nocite{seaton} have shown that a
 detailed calculation using the methods of the Opacity Project (OP,
 see Seaton et al. 1994)\nocite{opacityproject} for a six element
 mixture yields a Rosseland-mean opacity in the region of interest of
 order 5\% larger than the OPAL opacity for the same mixture.

All the models assume a solar age of $4.57\times 10^9$~yr,  a
present solar luminosity  ${\rm  L}_\odot=3.8418\times
10^{33}$~ergs$^{-1}$,   and a present solar radius of R$_\odot=
6.9598 \times 10^{10}$~cm. For each model, OPAL opacity tables
were used that correspond to the detailed composition that was
adopted.

\section{HELIOSEISMOLOGICAL INVERSIONS}
\label{sec:inversions}

We summarize in this section the largely standard techniques that
we use to determine the differences between the solar model
characteristics and the properties of the Sun as determined by
helioseismological measurements.

Helioseismological inversions generally proceed through a
linearization of the equations of stellar oscillation, using their
variational property, around a known reference model (see, e.g.,
Dziembowski, Pamyatnykh \& Sienkiewicz 1990\nocite{dziem90};
D\"appen et al.~1991\nocite{da91}; Antia \& Basu
1994\nocite{hmasb94}; Dziembowski et al.~1994;\nocite{dziem94}
 Elliott 1995; \nocite{elliott} Tripathy and Christensen-Dalsgaard 1998\nocite{tripathy2}).
 The differences between the
structure of the Sun and the reference model are then related to
the differences in the measured oscillation frequencies of the Sun
and the model by known kernels. Non-adiabatic effects and other
errors in modeling the surface layers give rise to frequency
shifts which are not accounted for by the variational principle
(Cox \& Kidman 1984\nocite{cox84}; Balmforth
1992)\nocite{balmforth92b}.  Since the eigenfunctions of low- and
medium-degree modes are essentially independent of degree in the
near-surface layers, the frequency shifts are just a function of
mode frequency, divided by the mode inertia (Christensen-Dalsgaard
\& Berthomieu 1991\nocite{jcd91}; Christensen-Dalsgaard \&
Thompson 1997\nocite{jcd97}). The frequency of a deeply
penetrating mode is shifted less by near-surface perturbations
than that of a shallowly penetrating mode of the same frequency.
In the absence of any first-principle formulation, these effects
are usually taken into account in an {\it ad hoc} manner by
including an arbitrary function of frequency in the variational
formulation (Dziembowski et al.~1990)\nocite{dziem90}. Thus, the
fractional change in the frequency of a mode can be expressed in
terms of the fractional changes in the structure of the model,
which can be characterized, for example, by the adiabatic sound
speed, $c$, and the density, $\rho$, as well as a surface term.
After linearization, one obtains:
\begin{equation}
{\delta \nu_i \over \nu_i}
=  \int_0^{R_\odot} K_{c^2,\rho}^i(r){ \delta c^2(r) \over c^2(r)}d r +
 \int_0^{R_\odot} K_{\rho,c^2}^i(r) {\delta \rho(r)\over \rho(r)} d r
 +{F_{\rm surf}(\nu_i)\over I_i}
\label{eq:inv}
\end{equation}
(e.g., Dziembowski et al.~1990)\nocite{dziem90}.  Here $\delta \nu_i$ is the
difference in the frequency $\nu_i$ of the $i$th mode between the
solar data and a reference model, $i$ representing the pair
$(n,l)$, where $n$ is the radial order and $l$ the degree of the
model. The kernels $K_{c^2, \rho}^i$ and $K_{\rho, c^2}^i$ are
known functions of the reference model which relate the changes in
frequency to the changes in $c^2$ and $\rho$, respectively. The
term involving $F_{\rm surf}$ takes into account the near-surface
errors in modeling the structure and the modes, and $I_i$ is the
mode inertia of the $i^{\rm th}$ mode.

Equation~(\ref{eq:inv}) constitutes the inverse problem that must be
solved to infer the differences in structure between the Sun and the
reference model. The inversions shown in this paper have been carried
out using the the Subtractive Optimally Localized Averages (SOLA)
technique (Pijpers \& Thompson 1992\nocite{fpp92},
1994)\nocite{fpp94}.  Details of how SOLA inversions are carried out
and how various parameters of the inversion are selected are given by
Rabello-Soares, Basu, \& Christensen-Dalsgaard (1999)\nocite{ra99a}.

In this paper, we use helioseismic inversions to determine how
similar the different solar models discussed in
\S~\ref{sec:models} are to the real Sun. Each of the models
described in \S~\ref{sec:models} is used as a reference model. For
the helioseismological data, we use solar oscillation frequencies
obtained by the Michelson Doppler Imager (MDI) on board the Solar
and Heliospheric Observatory (SOHO). In particular, we use
frequencies obtained from MDI data that were collected for the
first 360 days of its observation (Schou et
al.~1998)\nocite{sch98}. This data set was chosen because it was
derived from a long time series when solar activity was low. The
length of the time series results in reduced noise, and hence a
larger number of modes for which the frequencies can be determined
reliably. Mode-sets derived from longer data sets are available,
but they only consist of low degree modes (e.g., Bertello et
al.~2000)\nocite{ber00}. Also, a longer time series would have
meant adding observations from periods of increasing solar
activity, which would have changed the frequencies. It is a well
established fact that solar frequencies increase with solar
activity.  However, it is also known that the increase occurs as
an increase in the surface term in Eq.~\ref{eq:inv}, and hence
does not change inversion results (Basu 2002)\nocite{basu02}.

We invert for both the sound-speed differences and the density
differences between the solar models and the Sun.

\section{COMPARISONS BETWEEN SOLAR MODELS AND  OBSERVATIONS}
\label{sec:comparison}

In this section, we compare solar parameters determined from
helioseismological measurements with the values obtained from the
six solar models that are discussed in \S~\ref{sec:models}.
Table~\ref{tab:comparisons} summarizes the principal results.

\begin{figure}[!t]
\begin{center}
\includegraphics[bb= 70 70 550 400, width= 12cm]{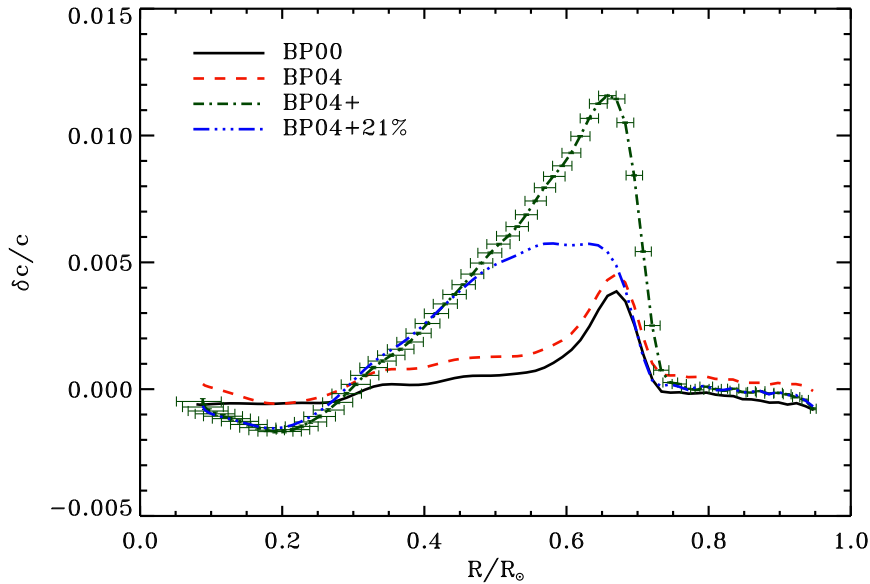}
\end{center}
\caption{Relative  sound-speed  differences,  $\delta  c/c=(c_\odot  -  c_{\rm
    model})/c_{\rm  model}$,  between   solar  models  and  helioseismological
    results from MDI data.  The vertical error bars show the $1\sigma$ error in the
inversion due to statistical errors in the data. The horizontal
error bars are a measure of the resolution of the inversions,
defined as the distance between the first and third quartile
points of the averaging kernels (approximately the half-width in
radius of the measurement in regions of good resolution).
\label{fig:velocitydiffs}}
\end{figure}

\begin{figure}[t]
\begin{center}
\includegraphics[bb= 70 70 550 400, width=12cm]{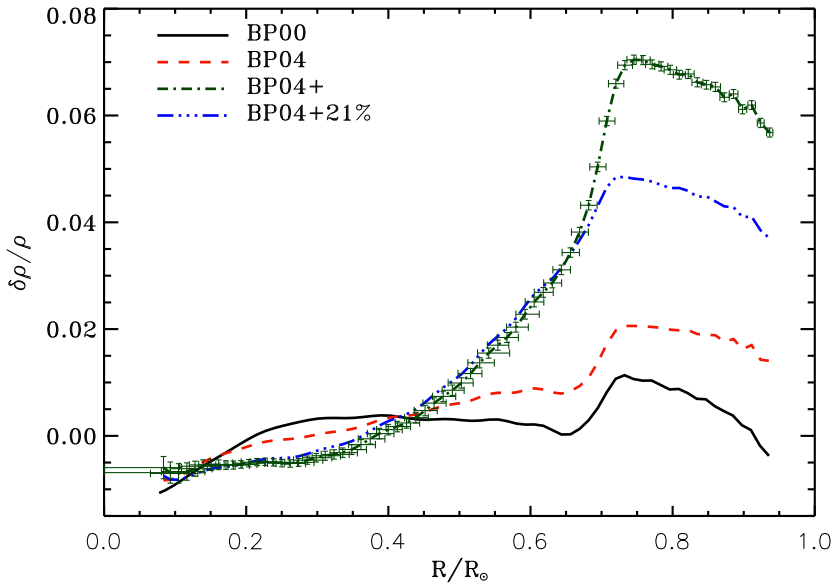}
\end{center}
\caption{Relative   density  differences,   $\delta   \rho/\rho=(\rho_\odot  -
  \rho_{\rm    model})/\rho_{\rm   model}$,    between   solar    models   and
  helioseismological results from MDI data.  The vertical error bars show the $1\sigma$ error in the
inversion due to statistical errors in the data. The horizontal
error bars are a measure of the resolution of the inversions,
defined as the distance between the first and third quartile
points of the averaging kernels (approximately the half-width in
radius of the measurement in regions of good resolution). For the
density, the resolution near the center of the Sun is particularly
poor.}\label{fig:densitydiffs}
\end{figure}

Figure~\ref{fig:velocitydiffs} shows the fractional differences
between the sound speeds as a function of solar radius that are
computed for each of the solar models and the sound speeds
determined from helioseismology. Figure~\ref{fig:densitydiffs}
shows for the density profiles a similar trend as
Figure~\ref{fig:velocitydiffs} shows for the sound speeds.  Since
it is well known that sound speed determinations are more accurate
and more robust than density determinations, we do not discuss
further the density profiles other than to remark that they are
consistent with all of the other comparisons we make between solar
model predictions and helioseismological measurements.

\begin{table}[!t]
\begin{center}
\caption{Solar model predictions versus helioseismological
determinations. The table presents for comparison with
helioseismological measurements the results of  a series of four
solar models discussed in \S~\ref{sec:models}. The successive
columns give the model designation, the adopted present  heavy
element to hydrogen mass ratio at the solar surface, the rms
fractional difference between the solar model sound speeds and the
helioseismologically-determined sound speeds, the rms fractional
difference for the density, the radius of the convective zone, and
the present surface helium abundance. For consistency, all the
results reported in this table were obtained with the Garching
stellar evolution code. \label{tab:comparisons}}
\smallskip
\begin{tabular}{lccccc}
\hline \hline
\noalign{\smallskip}
 MODEL & $Z/X$ &$\sqrt{<\left( c - c_\odot
\right)^2/c^2>} $ &$\sqrt{<\left( \rho - \rho_\odot
\right)^2/\rho^2>} $& $R_{\rm CZ}/{\rm R}_\odot$
& $Y_{\rm surf}$  \\
\hline
BP00 & 0.0229  & 0.0010 & 0.005& 0.7141 & 0.243 \\
BP04  & 0.0229 & 0.0014 & 0.011 & 0.7146  & 0.243\\
BP04--EOS96 & 0.0229 & 0.0013 & 0.012 & 0.7148 & 0.243 \\
BP04+ & 0.0176   & 0.0046   &  0.037 & 0.7259   & 0.238
 \\
BP04+ 21\% & 0.0176 & 0.0029 & 0.027 & 0.7133 & 0.239  \\
BP04+ 11\% & 0.0176 & 0.0014 & 0.013 & 0.7162 & 0.243 \\
\hline
\end{tabular}
\end{center}
\end{table}

\subsection{Comparisons for models BP00 and BP04: 1998 element abundances}
\label{subsec:bp00bp04}

The third column of Table~\ref{tab:comparisons} presents the
fractional rms differences between each solar model (used as a
reference model, see \S~\ref{sec:inversions}) and the
helioseismologically determined sound speeds.  We see that the
BP00 and the BP04 solar models, both of which are computed using
the older Grevesse \& Sauval (1998)\nocite{oldcomp} heavy element abundances, are
in good agreement with the solar sound speeds.  The rms agreement
with the solar sound speeds is about 0.1\% for both BP00 and BP04.
Figure~\ref{fig:velocitydiffs} shows the agreement between the
sound speeds predicted by the BP00 solar model (dark line) and the
BP04 solar model (dashed line).

The fifth column of Table~\ref{tab:comparisons} shows that  the
calculated depth of the convective zone for the BP00 and the BP04
models is in satisfactory agreement with the the measured value of
the depth of the convective zone given in
equation(\ref{eq:radiuscz}), $0.713R_\odot$.

The surface helium abundance of the Sun has recently been
redetermined by Basu and Antia (2004).\nocite{basu04}  They find
\begin{equation}
Y_{\rm helioseismology} = 0.2485 \pm 0.0034.
\label{eq:Yhelio}
\end{equation}
The interpretation of the error given in
equation~(\ref{eq:Yhelio}) is not simple since systematic
uncertainties are dominant. The fifth column of
Table~\ref{tab:comparisons} shows that the present-day surface
helium abundance obtained from models BP00 and BP04 may be
slightly lower than is obtained from helioseismology, but the
statistical significance of this difference is uncertain.

For completeness, we have computed a model that is identical to
BP04 except that instead of using the 2001 OPAL equation of state,
as was done in deriving the model BP04, we use the older 1996 OPAL
equation of state.  The results are given in the third row of
Table~\ref{tab:comparisons}.

Improvements in the equation of state between 1996 and 2001 are
reflected in  Table~\ref{tab:comparisons} by the slightly
different  values that are found for BP04 (row two)  and
BP04--EOS96 (row three). We see that the improvement in the
equation of state does not affect significantly the agreement of
the solar model results with the measured helioseismological
values. We conclude that plausible changes in the equation of
state are unlikely to explain the discrepancy between solar model
predictions and helioseismological measurements when the lower
metal abundances are used.

 As discussed in Basu, Pinsonneault, and Bahcall (2001), the effect of  mixing in the
radiative zone of the Sun would be in the direction to reconcile
the meteoritic and solar photospheric lithium abundances and to
bring the computed  surface helium slightly closer to the measured
value. Such models have a somewhat shallower solar surface
convection zone and the overall agreement with the sound speed
data is comparable, or slightly less good, than   models without
extra mixing.

\subsection{Comparisons for model BP04+ : new heavy element abundances}
\label{subsec:bp04+}

Figure~\ref{fig:velocitydiffs} shows the dramatic lack of
agreement between the helioseismological sound speeds and the
values predicted by the BP04+ solar model, which uses the new
heavy element abundance determinations that lead to $Z/X =
0.0176$. The biggest discrepancy is in the vicinity of the base of
the convective zone, near $0.7R_\odot$. However, there is a
significant discrepancy between BP04+ and the helioseismological
values all the way into about $0.3 R_\odot$.

Table~\ref{tab:comparisons} summarizes the magnitude of this
discrepancy. For the solar model BP04+, the rms discrepancy in the
sound speeds is more than a factor of three worse than for the
BP04 model (and more than a factor of five worse than for the BP00
model).  Furthermore, the depth of the convective zone, $0.726
R_\odot$, given in column~6 of Table~\ref{tab:comparisons} is
inconsistent with the measured value of $0.713 R_\odot$.Finally,
the surface helium abundance given in column~6, $Y = 0.238$, is
lower than the measured value given in equation~(\ref{eq:Yhelio}).

We conclude that the solar model BP04+, which is constructed using
the most recent heavy element abundance estimates, is inconsistent
with helioseismological measurements.

\subsection{Comparisons for BP04+  21\%: enhanced opacity  new abundances}
\label{subsec:bp04+21}

The comparison between the predictions of the model BP04+  21\%
and the helioseismological data is very instructive. This solar
model was investigated in Bahcall, Serenelli, \& Pinsonneault
(2004)\nocite{BSP04} because the 21\% increase in the radiative
opacity relative to the standard OPAL opacity was found to be
sufficient to resolve the discrepancy in the calculated depth of
the convective zone that was obtained with BP04+ model (with no
enhanced opacity). For a related discussion, see the paper by
Tripathy, Basu, and Christensen-Dalsgaard 1998\nocite{tripathy1}.

The  BP04+  21\% model was constructed with exactly the same input
data as for the BP04+ model, including the recent heavy element
abundance determinations, but in addition BP04+  21\% has the
radiative opacity increased artificially  by 21\% near the base of
the convective zone. The precise form of the opacity increase was
postulated to be of the form obtained by multiplying the OPAL
opacity in the vicinity of the convective envelope boundary by a
Lorentzian function $f(T)$.  Specifically, the multiplicative
factor $f(T)$ was taken to be
\begin{equation}
f(T)= 1 + \frac{\alpha \gamma^2}{\left( (T-T_0)^2 + \gamma^2
\right)} \, . \label{eq:defnofperturbation}
\end{equation}
Here $T$ is the temperature in the  solar model. The perturbed
opacity is  $\kappa_{\rm perturbed} = \kappa_0 f(T)$, where
$\kappa_0$ is the unperturbed radiative opacity, $\alpha$ is the
amplitude of the perturbation, and $\gamma$ is the width of the
perturbation (defined as the point where the perturbation drops to
$\alpha /2$). The temperature at the base of the CZ is $T\approx
2.18\times 10^6$K,  which was used for $T_0$ in
equation~(\ref{eq:defnofperturbation}). The BP04+  21\%  solar
model was calculated  for a width of the opacity perturbation
$\gamma= 0.2\times 10^6 {\rm \, K} \approx 0.1T_0$. This value of
$\gamma$ corresponds to a width in the solar radius of only
$\Delta R = 0.02 R_\odot$.

Figure~\ref{fig:velocitydiffs} shows two things about the
\hbox{BP04+ 21\%} solar model.  First, the 21\% increase in the
opacity near the base of the solar convective zone indeed improves
significantly the agreement with the measured sound speeds over
what is obtained with the model BP04+. Second, the improved
agreement is limited to the region near the base of the convective
zone and there remains a significant disagreement down to radii of
order  $0.4 R_\odot$ ($T = 5\times10^6$K). Of course,  different
assumed forms of the factor $f(T)$ lead to different estimates of
how much opacity change is required to construct a model with the
correct depth of the convection zone (see, e.g., Basu \& Antia
2004).\nocite{basu04}

In summary, Figure~\ref{fig:velocitydiffs} indicates that the
radiative opacity would have to  be changed in a broad range of
temperatures (radii) in order to resolve the discrepancies between
helioseismological measurements and solar model predictions made
using the new heavy element abundances. The relatively low value
for the surface helium abundance, $Y = 0.239$ obtained with
\hbox{BP04+ 21\%} (see Table~\ref{tab:comparisons}), may also
reflect the need for an opacity correction that extends down to
$\sim 5 \times 10^6$K.

\subsection{Comparisons for BP04+  11\%}
\label{subsec:bp04+11}

\begin{figure}[t]
\begin{center}
\includegraphics[angle= 90, width=17cm]{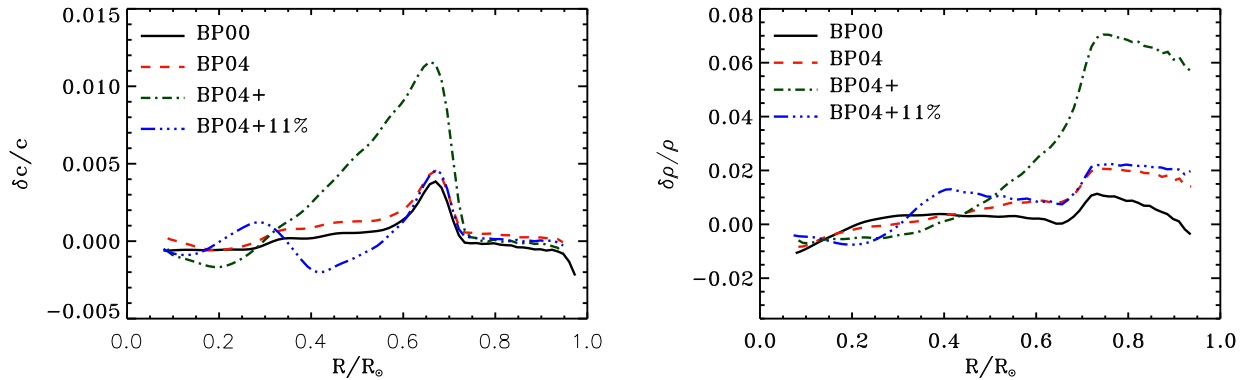}
\end{center}
\caption{ A model with an 11\% opacity increase. The figure shows
the relative sound speed and density differences (left and right
panels
  respectively) between  solar models and helioseismological  results from MDI
  data. The BP04+11\% model has  the same characteristics as BP04+ (i. e., low metal abundances)
  except that the radiative opacities have been increased by a constant 11\%
factor from the base of the convective zone down  to $5 \times
10^6$K ($R = 0.4 R_\odot $).}\label{fig:diffsbp04+11}
\end{figure}

Motivated by  the results of \S~\ref{subsec:bp04+21}, we have
computed a variety of solar models assuming the correctness of the
recently determined  low metal abundances but with different
assumed opacity changes. We have studied the helioseismological
properties of these models. The reader will immediately recognize
that one can in principle consider an infinite number of such
`low-metal, higher-opacity' models, with prescriptions for
changing the opacity of varying complexity and artificiality . We
acknowledge that there is limited utility in computing such models
without a physical basis for the assumed opacity changes.

However, we have found a relatively simple prescription for
changing the opacity, while adopting the low metal abundances,
that yields reasonable agreement with the observed
helioseismological properties. We present the results for this
model here not out of any conviction that the assumed opacity law
is correct, but rather to illustrate the general quality of the
fit to the helioseismological data that is possible and to
indicate approximately how much the opacity would have to be
shifted in order to obtain reasonably good agreement with the
helioseismological measurements.

The results of \S~\ref{subsec:bp04+21} indicate that the opacity
must be changed over a relatively broad range of temperatures if
we adopt the lower metal abundances. For simplicity, we assumed a
constant 11\% increase above the OPAL opacity from $2\times 10^6$K
($R = 0.7R_\odot$) down to $5\times 10^6$K ($R = 0.4 R_\odot$),
where the opacity increase was smoothly turned off (half-width of
turn off is $2 \times 10^5$K). We denote this model by BP04+11\%.

We are sure that the  prescription of a constant opacity increase
that is implemented in BP04+11\% is too simple to represent the
improvements in the radiative opacity that are likely to result
from detailed quantum mechanical calculations of the solar mixture
of hydrogen, helium, and heavy elements. But, we shall see that
this model with a constant opacity increase fits the data
reasonably well and is a crude approximation to what might guess
is required by comparing (see Figure~\ref{fig:opacitydifferences}
below) the opacities in the BP04 model (successful in describing
the helioseismological data) and the BP04+ model (unsuccessful in
describing the helioseismological data).

Figure~\ref{fig:diffsbp04+11} and Table~\ref{tab:comparisons} show
that the BP04+11\% solar model fits the helioseismological data
with an accuracy that is comparable to our best-fitting solar
models, BP00 and BP04.  We conclude that an increase in the
opacity of the order of 10\% in the range $2\times 10^6$K to
$5\times 10^6$K would resolve the discrepancy between the
predictions of solar models computed with the new lower metal
abundances and the helioseismological measurements.

\begin{figure}[t]
\begin{center}
\includegraphics[width=12cm]{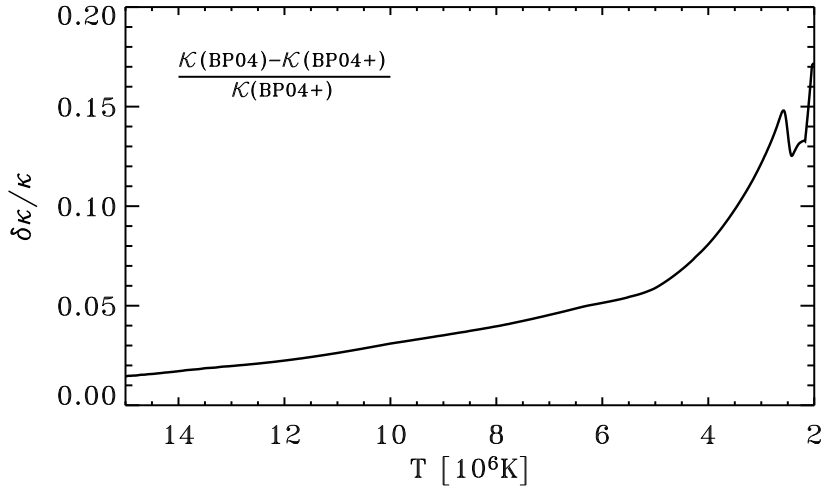}
\end{center}
\caption{ Opacity difference between BP04 and BP04+ solar models.
 The figure shows the fractional opacity difference between the
  two solar models BP04~(higher metal abundances) and BP04+~(lower metal abundances)
  as a function of the temperature in the BP04 solar model.
  }\label{fig:opacitydifferences}
\end{figure}

Figure~\ref{fig:opacitydifferences} shows how the OPAL radiative
opacity changes due  to the adoption  of the new solar
composition.We have evaluated the opacity at the same temperatures
and densities for two solar models, BP04 and BP04+, that differ
only in their assumed composition.  To correct for the small
effect that at the same temperature the density differs slightly
in the two models, we use the following equation:
\begin{equation}
\left(\frac{\delta \kappa}{\kappa}\right)_{T,\rho} ~\equiv~
\frac{\kappa_{04}(T,\rho)-\kappa_{04+}(T,\rho)}{\kappa_{04+}(T,\rho)},
\label{eq:deltakappacomposition}
\end{equation}
where
\begin{equation}
\kappa_{04+}(T,\rho) \approx \kappa_{04+}(T,\rho') +
\left(\frac{\partial \kappa_{04+}}{\partial \rho'}\right)_T (\rho
- \rho') . \label{eq:kappataylorseries}
\end{equation}
Here $\kappa_{04}$  and $\kappa_{04+}$ are the opacities
corresponding to the BP04 and BP04+ solar models respectively, $T$
and $\rho$ are temperatures and densities at a given point in the
BP04 model and $\rho'$ is the density in the BP04+ model at the
temperature $T$. Including the density dependence, makes very
little difference near the base of the CZ but increases the
fractional opacity difference by about 40\% of its value at the
highest temperature ($T = 5 \times 10^6$ K) at which an opacity
perturbation was introduced into the solar model BP04+11\%. The
fractional difference is small (less than 3\%) in the regions
where  solar neutrinos are produced  ($R < 0.2 R_\odot$  and $T  <
9\times 10^6$~K). However,  for  $T<5\times 10^6$~K  the
difference increases  and  reaches  about  15\% close to  the base
of  the convective zone. Figure~\ref{fig:opacitydifferences} shows
why the BP04+ 11\%  model approximately restores  the agreement
with helioseismological measurements.

We have evolved solar models with larger opacity increases near
the base of the convective zone and smaller increases further in,
but we have not obtained by this procedure a substantial
improvement in the agreement with helioseismological data over
what is achieved with the BP04+11\% model.

Basu \& Antia (2004) discussed the characteristics of a solar
envelope model which invoked, in order to satisfy helioseismic
constraints, a 19\% increase in radiative opacity relative to the
tabulated OPAL opacities. We have evolved a full solar model with
the same opacity increase and heavy element abundance as the Basu
\& Antia (2004) model (i.e. a 19\% increase in opacity from the
base of the convection zone to a temperature of $5\times10^6$K and
the heavy element to hydrogen mass ratio of Z/X=0.0171). We find,
as expected from the previous discussion of BP04+11\% and from
Figure~\ref{fig:diffsbp04+11}, that the 19\% increase in opacity
is too large to provide a good fit to the helioseismological data.
The depth of the convection zone for the evolved model is
$R=0.708R_\odot$ and the rms fractional sound speed discrepancy is
$\delta c/c = 0.0033$. The reason for the difference in our
conclusion and the Basu \& Antia (2004) result almost certainly
lies in the fact that the Basu \& Antia envelope model was forced
to have abundance profiles near the base of the convective zone
that are different from what we find in our stellar evolution
models, while at the same time being  silent about the
helioseismological properties in the radiative interior. The Basu
\& Antia envelope model was forced to have  heavy-element and
helium profiles in agreement with the helioseismological
determinations near the base of the convective zone. For standard
solar  models, the heavy element and helium profiles are different
from that of the Sun near the base of the convective zone (Basu \&
Antia 1994;~\nocite{basuantia94} Bahcall, Pinsonneault,
  Basu, \& Christensen-Dalsgaard 1997;\nocite{bpbc97} Antia \& Chitre
1998~\nocite{antiachitre98}), probably because of turbulent mixing
not included in the standard models (Elliott \& Gough
1998\nocite{elliottgough98}). Over the radiative interior of the
Sun, $R = 0.0$ to $R = 0.7 R_\odot $ standard solar models like
BP00 or BP04 are, as we have seen, in excellent agreement with the
helioseismological data.


\section{CONCLUSIONS}
\label{sec:conclusions}

We summarize and discuss in this section our five principal
conclusions. The main quantitative results of our studies are
given in Table~\ref{tab:comparisons} and in
Figure~\ref{fig:velocitydiffs} and Figure~\ref{fig:diffsbp04+11}.

\begin{description}
\item[ (1) Larger heavy element abundances yield satisfactory
solar models.] Standard solar models constructed with the older
(i. e., higher) heavy element abundances (models BP00, BP04, and
BP04-EOS96) are in good agreement with the helioseismological
data.  The solar sound speeds, depth of the convective zone, and
surface abundance of helium determined from helioseismology are
all in agreement with the values obtained from these solar models
that were computed using the older element abundances. We
interpret the differences between the predictions of the models
BP00, BP04, and BP04--EOS96 as indicating the expected range of
characteristic parameters that can occur with typical improvements
in the input data to the solar models.

\item[(2) Standard models with less heavy elements disagree with
helioseismology.] A solar model constructed with the new heavy
element abundances, BP04+, is inconsistent with the
helioseismologically measured sound speeds, the depth of the
convective zone, and the surface helium abundance.

\item[ (3) Increasing the opacity near the base of the CZ helps,
but is not enough.] We calculate a solar model using the newer
heavy element abundances   and also increase the radiative opacity
near the base of the solar convective zone by just the amount
required to make the CZ depth calculated with new heavy element
abundances agree with the measured depth. The improved agreement
between the solar model and the helioseismological determinations
is limited, like the assumed change in the radiative opacity, to
regions near the base of the convective zone.

\item[ (4) A 11\% increase in opacity over a broader range is
okay.] Suppose that a change in the OPAL radiative opacity is
required to explain the reason why solar models constructed with
the newer heavy element abundances are in conflict with
helioseismology measurements. Then the OPAL opacity  must be
increased by about 11\% from about $2.2\times 10^6$K at the base
of the CZ ($R = 0.71R_\odot$) all the way down to about $5 \times
10^6$K ($R = 0.4 R_\odot$). It would be very useful to study
whether such a change in opacities is consistent with other
astronomical data. The required 11\% increase is larger than the
difference reported by Seaton  \& Badnell (2004) between the
radiative opacities calculated independently by the Opacity
Project and by the OPAL project.

\item[ (5) The predicted solar neutrino fluxes are not
significantly affected. ]  The differences in the predicted solar
neutrino fluxes for the most different solar models considered in
this paper, BP04 and BP04+, are all within the $1\sigma$ quoted
theoretical errors (see Table~1 of Bahcall and Pinsonneault 2004).
If we compare models that differ only in whether or not a 11\%
increase in opacity has been included, the differences in
predicted neutrino fluxes are slightly smaller, especially for the
most important neutrino sources: 1\% ($p-p$ neutrinos), 2\%
($^7$Be neutrinos), and 6\% ($^8$B neutrinos).

\end{description}

There are, in addition to the opacity, other sources of potential
change in the solar model input data, most importantly the
uncertainties in the measurements of the heavy element abundance
and the uncertainties in the calculation of the heavy element
diffusion coefficients. The recent heavy element abundance
determinations have quoted uncertainties of order 0.05 dex (12\%)
(see Asplund et al.~2000; \nocite{asplundetal00} Asplund
2000;\nocite{asplund00} Allende Prieto, Lambert, \& Asplund
2001;\nocite{allende01} Allende Prieto, Lambert, \& Asplund
2002;\nocite{allende02} Asplund et al.~2004\nocite{asplund04}).
The heavy element diffusion coefficients are uncertain by about
15\% (see Thoul, Bahcall, and Loeb 1994).

It may well be that the correct reconciliation of abundance
determinations will involve modest adjustments relative to the
present standard values of all of the factors mentioned above,
namely, the abundances themselves, the diffusion coefficients, and
the radiative opacity. The increase of the radiative opacity by
11\% obtained in this paper with the help of the model BP04+ 11\%
may be regarded as a plausible upper limit to the opacity
correction that is required since it assumes no change in any of
the other input parameters.

Why have we not constructed and explored even more solar models
with a variety of hypothetical changes in the radiative opacity,
diffusion coefficients, and heavy element abundances? The reason
is that  for the opacity changes by themselves there is an
infinity of conceivable corrections, with different amplitudes and
shapes. Moreover, one can assume whatever changes one wants,
within the quoted uncertainties, for the diffusion coefficients
and the heavy element abundances. Improved calculations of the
radiative opacity (see, \hbox{e. g.,} Seaton and Badnell
2004\nocite{seaton} for recent refinements) will determine what,
if any, significant refinements are implied by more accurate
calculations. Once those calculations are available it will be
appropriate to make new solar models to incorporate the newly
calculated opacities.

\acknowledgments J. N. B. is supported in part by NSF grant
PHY-0070928. S. B. was partially supported by NSF grants ATM
0206130 and ATM 0348837. A. M. S is supported in by the W. M. Keck
Foundation through a grant to the Institute for Advanced Study. We
are grateful to M. Seaton for stimulating comments.

This work utilizes data from the Solar Oscillations
Investigation/Michelson Doppler Imager (SOI/MDI) on the Solar and
Heliospheric Observatory (SOHO).  SOHO is a project of
international cooperation between ESA and NASA. MDI is supported
by NASA  contract NAG5-13261 to Stanford University.

\end{document}